\documentclass[useAMS,usenatbib]{mn2e}

\usepackage{lscape,graphicx,ulem}

\title[On the difference between Herbig Ae and Be stars]
  {On the difference between Herbig Ae and Herbig Be stars}
\author[J.C. Mottram et al.]
  {J.C. Mottram$^1$\thanks{E-mail:jcm@ast.leeds.ac.uk}, J.S. Vink $^{2,3}$, R.D. Oudmaijer $^1$, M. Patel $^{1,3}$
\\
  $^1$Department of Physics and Astronomy, University of Leeds, Leeds LS2 9JT, UK\\
  $^2$Lennard-Jones Laboratory, Astrophysics, Keele University, Keele ST5 5BG, UK\\
  $^3$Imperial College London, Blackett Laboratory, Prince Consort Road, London SW7 2AZ}

\begin{document}

\date{Accepted 2007 March 3. Received 2007 January 5; in original form 2006 October 4}

\pagerange{\pageref{firstpage}--\pageref{lastpage}} \pubyear{2006}

\maketitle

\label{firstpage}

\begin{abstract}

We present linear spectropolarimetric data for eight Herbig Be and four Herbig Ae stars at 
H$\alpha$, H$\beta$ and H$\gamma$. Changes in the linear polarisation are detected across 
all Balmer lines for a large fraction of the observed objects, confirming that the small-scale regions 
surrounding these objects are flattened (i.e. disk-like). 
Furthermore, all objects with detections show similar characteristics at the three spectral lines, despite 
differences in transition probability and optical depth going from H$\alpha$ to H$\gamma$. 
A large fraction of early Herbig Be stars (B0-B3) observed show line depolarisation effects. However the 
early Herbig Ae stars (A0-A2), observed for comparison, show intrinsic line 
polarisation signatures. Our data suggest that the popular magnetic accretion scenario for T Tauri objects 
may be extended to Herbig Ae stars, but that  it may not be extended to early Herbig Be stars, for which
the available data are consistent with disc accretion. 
 
\end{abstract}


\section{Introduction}

While magnetic T-Tauri type models for low-mass star formation have
been reasonably well accepted, there is currently no such picture for
the early evolutionary phases of higher mass stars.  Many questions,
such as whether higher mass stars form via disc accretion, and issues relating to the importance
of magnetic fields during high-mass star formation, remain open. Herbig Ae/Be
stars are of considerable interest in attempts to resolve these
questions, as they are young stars of intermediate mass (2 -
15M$_{\odot}$) at the interface between low and high-mass star
formation.  In addition, Herbig Ae/Be stars are the most massive stars
with a visible pre-main sequence phase.

T Tauri stars have relatively strong magnetic fields \citep{b6,b8} and
these fields have important physical effects on the accretion flow.
It is generally believed that magnetic fields play a lesser role for
stars more massive than the Sun, as stars at spectral types A and earlier
lack convective outer mantles.

The picture of the role of magnetic accretion as a function of
spectral type changed when \citet{b2} found that a large
fraction (9/11) of Herbig Ae stars show changes in the linear
polarisation intrinsic to the H$\alpha$ line \citep[extended to 10/13 with data from ][]{b3} 
-- similar to the
fraction (9/10) of such intrinsic line effects in T Tauri stars \citep{b3}.  The origin for these H$\alpha$ line
effects in both T Tauri and Herbig Ae stars is believed to be the result of
the fact that {\it emission line} photons from the central object are scattered
off a rotating accretion disc with an inner hole \citep{b22}. From here the material is probably 
funnelled along magnetic field lines \citep{b16,b17}, onto the stellar surface.

\begin{table*}
\centering
\caption{Herbig Ae/Be observations. Columns 2~\&~3 show the logarithmic average of the two photometric magnitude measurements in the relevant band by \citet{b21}. 
Spectral types (column 4) are taken from references in \citet{b18}. 
The exposure time (columns 7~\&~8) consists of an integer multiple of 4 
spectra (one for each half-wave plate position) multiplied by the exposure time of an individual spectrum.}
\begin{tabular}{@{}cccccccc@{}}
\hline
\centering
Object Name & \multicolumn{2}{c}{Mag.} & Sp.~Type & \multicolumn{2}{c}{Date} & \multicolumn{2}{c}{Exposure(s)}\\
& $R$ & $B$ & & $R$ & $B$ & $R$ & $B$ \\
\hline
MWC 166 &6.90&7.35& B0 IV & 30/12/96 + 01/01/97$^{\bullet}$ & 29/09/04 &$-$& 480  \\
MWC 1080 &11.02&13.28& B0 &$-$& 28/09/04 &$-$& 4320 \\
&&&& 29/09/04 & 29/09/04 & 2880 & 3600  \\
GU CMa &6.64&6.67& B2 V & 11/07/95$^{\bullet}$ & 28/09/04 &$-$& 720 \\
MWC 361 &7.27&7.69& B2V & 18/12/99$^{*}$ & 28/09/04 &$-$ & 1600  \\
Il Cep &9.03&9.80& B2 IV$-$V & 19/12/99$^{*}$ & 29/09/04 &$-$& 1440 \\
BD +404124&10.53&11.37& B2 & 19/12/99$^{*}$ & 28/09/04  &$-$& 2160 \\
$\omega$ Ori &4.56&4.38& B3 III$-$IV & 11/01/95$^{\bullet}$ & 28/09/04 &$-$& 144 \\
MWC 147 &8.73&9.06& B6 V & 18/12/99$^{*}$  & 28/09/04&$-$ & 1920 \\
\hline
AB Aur &7.03&7.15& A0 V & 18/12/99$^{*}$ & 30/09/04 &$-$& 1920 \\
MWC 120 &7.92&7.95& A0 & 29/09/04 & 29/09/04 & 1080 & 2640 \\
MWC 480 &7.63&7.91& A2 & 30/09/04 & 30/09/04 & 1800 & 1320 \\
XY Per &9.41&10.04& A2 II & 20/12/99$^{*}$ & 29/09/04 &$-$& 1200 \\
\hline 
\multicolumn{8}{l}{$^{*}$ Data from \citet{b2}, $^{\bullet}$ data from \citet{b1}} \\
\label{tab:observations}
\end{tabular}
\end{table*}

Early Herbig Be stars \citep[spectral types B0-B5, see][]{b5} show
very different behaviour in their H$\alpha$ polarimetry. Although the
frequency of observed effects is high \citep[7/12; ][]{b2}, and wholly
consistent with all Herbig Be stars being embedded in flattened
circumstellar media (c.f. classical Be stars), their H$\alpha$
polarisation behaviour is very different from that of Herbig Ae/T
Tauri stars, as it is not the line but the continuum that is
found to be polarised, with the H$\alpha$ line ``depolarised'' with
respect to the continuum \citep{b1}. So, in summary, the emission line photons
are intrinsically polarised compared to the continuum for Herbig Ae and T Tauri stars, 
but are depolarised relative to the continuum for Herbig Be stars.

The difference in the behaviour of the linear polarisation across
H$\alpha$ between the Herbig Be and Ae stars may be an indication that
there is a transition in the Hertzsprung-Russell Diagram from
magnetic accretion at spectral type A to disc accretion at spectral
type B. However, alternatively, \citet{b2} considered the option that
the compact polarised H$\alpha$ emission occurring in the Herbig Ae
and T Tauri stars may be masked in the Herbig Be stars due to their much
higher levels of H$\alpha$ emission.

We therefore designed an observational test to distinguish between
these two options. The test involves linear spectropolarimetry across
less opaque, higher excitation emission lines that can be expected to
arise in the same location as the compact H$\alpha$ component that
causes the line polarisation in the Herbig Ae stars.  If the extension
and optical depth of H$\alpha$ emission masks the evidence
for a compact source in Herbig Be stars, then higher Balmer
(H$\beta$ and H$\gamma$) observations should reveal position angle
rotations ($QU$ loops) in the Herbig Be group. However, if these angle
rotations are not observed at these lines in the Herbig Be population,
while they do appear at H$\alpha$ among the Herbig Ae stars, this
would suggest that the circumstellar environments of Herbig Be stars are physically
different from those of Herbig Ae stars.  

The goal of this paper is thus to perform spectropolarimetry of
H$\beta$ and H$\gamma$ observations in early Herbig Be stars. The paper
proceeds as follows: we begin by discussing the observations and data reduction
 in $\S$2, then we discuss the results in $\S$3 before providing discussion of these results
and concluding remarks in $\S$4. 
Detailed comments of the results for each object in our sample are presented 
in Appendix A, divided between Herbig Be stars
($\S$~\ref{s_A_HBe}) and Herbig Ae stars ($\S$~\ref{s_A_HAe}) 
in our sample.

\section{Observations \& Data Reduction}

The linear spectropolarimetric data were taken using the 4.2m William
Herschel Telescope (WHT), La Palma over 3 nights in late September
2004.  The ISIS spectrograph, a calcite block, and a rotating
half-wave plate were used for the observations, as well as a dekker
and slit.  The dekker had three 5'' holes, separated by 18'',
while a 1'' slit width was used for all object observations .  The
calcite block acted to separate the incoming light into two
perpendicular rays, the O (ordinary) and E (extraordinary) rays, so
one image contains a set of O and E rays for the image and two sets
for the sky. Rotation of the half-wave plate then allowed  measurement of the polarisation at the angles 0$^\circ$,
45$^\circ$, 22.5$^\circ$ and 67.5$^\circ$. One complete data set is
therefore formed by combining four images, one at each of the four
angles mentioned above.

All bright early Herbig Be stars \citep[from the catalogue by ][]{b18} 
visible from La Palma in September 2004 were observed. We note that we strictly 
stick to their criteria, even
though the evolutionary status of objects like $\omega$~Ori is debated. 
These data were supplemented by four
Herbig Ae stars with similar criteria, as a basis for comparison.

\begin{table*}
\centering
\caption{Continuum polarisation measurements. Column 2 is the same as column 4 in Table $\ref{tab:observations}$. The continuum polarisation 
(columns 3~\&~4) and polarisation angle (columns 5~\&~6) were measured over the ranges 
(6150$-$6540~{\AA},6580$-$6815~{\AA}) and (4295$-$4310~{\AA},~4370$-$4840~{\AA},~4890$-$4955~{\AA}) for $R$ and $B$ respectively. 
Systematic errors in \%P and $\Theta$ are estimated at $\sim$0.1 per cent.}
\begin{tabular}{@{}cccccc@{}}
\hline
\centering
Object Name & Sp.~Type & \multicolumn{2}{c}{P$_{cont}$($\%$) } & \multicolumn{2}{c}{$\Theta$$_{cont}$($^\circ$) } \\
& & $R$ & $B$ & $R$ & $B$ \\
\hline
MWC 166 & B0 IV & 0.49 $\pm$ 0.01$^{\bullet}$ & 0.37 $\pm$ 0.01 & 44.9 $\pm$ 0.1$^{\bullet}$ & 37.2 $\pm$ 0.5 \\
MWC 1080 & B0 &$-$& 1.71 $\pm$ 0.02 &$-$&  72.2 $\pm$ 0.3 \\
& & 1.51 $\pm$ 0.01 & 1.99 $\pm$ 0.02 & 77.2 $\pm$ 0.1 & 72.1 $\pm$ 0.3 \\
GU CMa & B2 V & 1.15 $\pm$ 0.01 $^{\bullet}$& 1.25 $\pm$ 0.01 & 18.9 $\pm$ 0.1$^{\bullet}$ & 17.2 $\pm$ 0.1 \\
MWC 361 & B2V & 0.82 $\pm$ 0.01$^{*}$ & 0.88 $\pm$ 0.01 & 95.6 $\pm$ 0.1$^{*}$ & 96.4 $\pm$ 0.1 \\
Il Cep & B2 IV$-$V & 4.24 $\pm$ 0.01$^{*}$ & 4.18 $\pm$ 0.02 & 101.6 $\pm$ 0.1$^{*}$ & 100.7 $\pm$ 0.1 \\
BD +404124& B2 & 1.22 $\pm$ 0.01$^{*}$ & 1.37 $\pm$ 0.01 & 8.7 $\pm$ 0.2$^{*}$ & 17.4 $\pm$ 0.2 \\
$\omega$ Ori & B3 III$-$IV & 0.27 $\pm$ 0.01$^{\bullet}$ & 0.77 $\pm$ 0.01 & 55.7 $\pm$ 0.6$^{\bullet}$ & 43.9 $\pm$ 0.1 \\
MWC 147 & B6 V & 1.05 $\pm$ 0.01$^{*}$ & 0.85 $\pm$ 0.01 & 100.0 $\pm$ 0.1$^{*}$ & 102.4 $\pm$ 0.1 \\
\hline
AB Aur & A0 V & 0.11 $\pm$ 0.01$^{*}$ & 0.24 $\pm$ 0.01 & 53.3 $\pm$ 0.9$^{*}$ & 54.5 $\pm$ 0.6 \\
MWC 120 & A0 & 0.41 $\pm$ 0.01 & 0.40 $\pm$ 0.01 & 115.4 $\pm$ 0.2 & 117.0 $\pm$ 0.1 \\
MWC 480 1& A2 & 0.20 $\pm$ 0.01 & 0.33 $\pm$ 0.01 & 64.9 $\pm$ 1.3 & 64.3 $\pm$ 0.6 \\
XY Per & A2 II & 1.60 $\pm$ 0.01$^{*}$ & 1.60 $\pm$ 0.01 & 131.6 $\pm$ 0.1$^{*}$ & 125.5 $\pm$ 0.1 \\
\hline 
\multicolumn{6}{l}{$^{*}$ Data from \citet{b2}, $^{\bullet}$ data from \citet{b1}} \\
\label{tab:continuum}
\end{tabular}
\end{table*}

The data includes images in both the $B$ (4295~$-$4955~{\AA}) and $R$
(6150~$-$6815~{\AA}) bands, with mean wavelength dispersions of
0.22~{\AA}/pixel and 0.45~{\AA}/pixel respectively. For the $B$ band
data, the R1200B grating and the 4096x2048 pixel EEV12 CCD detector on
the ISIS Blue Arm were used, resulting in a spectral resolution
(measured from arc lines) of 51 km/s around H$\beta$. In the $R$ band,
observations were taken using the R1200R grating and the MARCONI2 CCD
on the ISIS Red Arm, resulting in a spectral resolution of 34 km/s
around H$\alpha$. In order to eliminate instrumental asymmetry,
observations for each object were taken with the object in the A (left
most) and B (central) dekker holes. The dekker hole C was not used for
object observations, but the sky was always measured at either A or B,
whichever was not being used for the target. Care was taken not to
saturate the CCD for objects with particularly strong emission lines
by taking short exposures. The seeing on the first two nights was
fair ($<1.7''$) but became poor ($>2''$) on the third night due
to clouds.

Data reduction was carried out on each image frame using the Figaro
software maintained by Starlink, and consisted of bias subtraction,
cosmic ray removal, bad pixel correction, spectrum straightening and
flat fielding. Wavelength calibration was performed using several
calibration spectra (obtained by observing a Copper-Argon-Neon lamp)
taken throughout the observing runs.  The images were then imported
into the Starlink program CCD2POL, which is part of the
Time-Series/Polarimetry (TSP) package, in order to obtain the Stokes
parameters for each polarisation set. The resultant data was then fed
into the POLMAP package also maintained by Starlink, in order to
measure the polarisation and polarisation angle, and to produce plots
of the data.

The objects observed included several polarisation and
zero-polarisation standards. These were observed so that the
instrumental polarisation and instrumental polarisation angle could be
measured by comparing the initial results for these objects with
the literature.  From these data, we found that the
polarisation P has a systematic error of $\simeq$ 0.1$\%$ and the
polarisation angle $\theta$ has a systematic error of $\leq$
1$^\circ$. While in principle the achieved accuracy is governed by
photon statistics alone as $\sqrt{N}$ (typically 0.01$\%$), it can be
seen that these calibration and telescope-induced errors are the main
limiting factor.

We do not correct for instrumental or interstellar polarisation
because these only add a wavelength-independent constant to the Stokes
$Q$ and $U$ parameters. Due to this wavelength independence, we
concentrate on the $(Q,U)$ representation of the hydrogen emission line
polarisation signatures when classifying the observed
behaviour. Furthermore, it is often not possible to distinguish the
effects of the interstellar medium from that of the material
immediately surrounding the target object such as dust.

\begin{figure*}
\centering
\includegraphics[width=60mm, height=140mm, angle=-90]{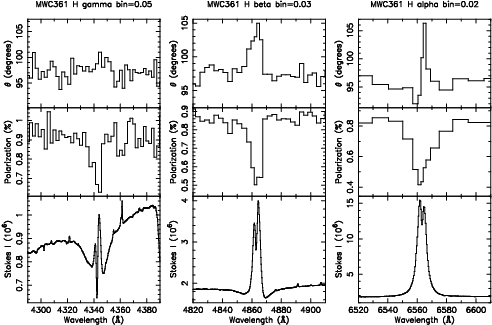}
\vfill
\vspace{2mm}
\includegraphics[width=40mm, height=140mm, angle=-90]{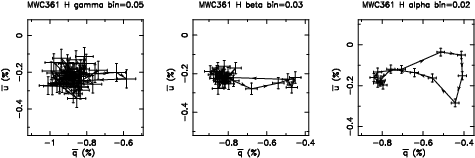}
\vspace{5mm}
\phantom{bliep} 
\noindent 
\line(1,0){475}
\vspace{5mm}
\centering
\includegraphics[width=60mm, height=140mm, angle=-90]{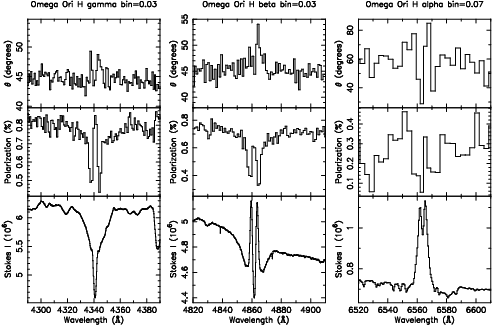}
\vfill
\vspace{2mm}
\includegraphics[width=40mm, height=140mm, angle=-90]{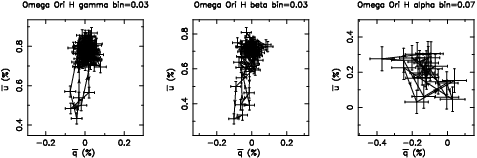}
\caption{The polarisation data of two Herbig Be stars. The H$\alpha$
data for MWC~361 is from 1999 and of $\omega$~Ori is from 1995. The
H$\beta$ and H$\gamma$ data of both objects were taken in 2004. Top figures: the
data are visualised using ``triplots'' (top) and {\it (Q,U)} diagrams
(bottom) of MWC~361 around H$\gamma$, H$\beta$ and H$\alpha$. In the
polarisation spectra, the bottom panel shows the Stokes intensity as a
function of wavelength, while the middle panel shows the percentage
polarisation and the upper plot the polarisation angle, also as
function of wavelength.  Bottom figures: as above, but now for
$\omega$~Ori. The difference between the $B$ and $R$ band continuum
data (see table $\ref{tab:continuum}$) suggests that 
the emission from the object is variable (see text).  Note that
the H$\gamma$ emission is hardly visible, but still yields a
polarisation signature consistent with that observed at H$\beta$.}
\label{F:Omega Ori}
\label{F:MWC361}
\end{figure*}

\begin{figure*}
\centering \includegraphics[width=60mm, height=140mm,angle=-90]{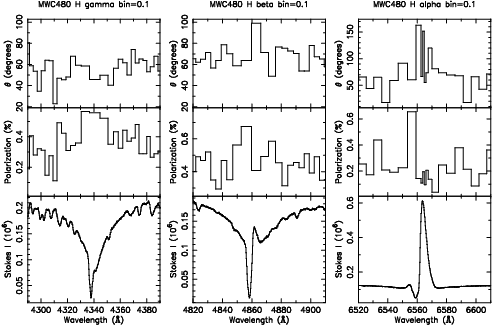} 
\vfill
\vspace{2mm}
\includegraphics[width=40mm, height=140mm, angle=-90]{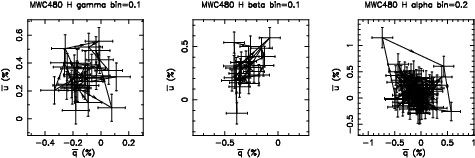}
\caption{As in Fig. $\ref{F:MWC361}$ but for the Herbig Ae star
MWC~480. The binning error for the {\it (Q,U)} plot for H$\alpha$ is
twice that of the triplot in order to provide a well-defined
continuum. All data is from 2004.}
\label{F:MWC480}
\end{figure*}

\section{Results}

We have listed the observational properties for both the $R$ and $B$
band observations for all Herbig Ae and Be stars in Table
$\ref{tab:observations}$. For 9 of the 12
objects, additional $R$ band data from \citet{b2} and \citet{b1} have
been listed. For 3 objects, H$\alpha$ data were taken in
2004. The continuum polarisation was measured for all targets
in both $B$ and $R$, and is listed in Table $\ref{tab:continuum}$. 
This acts as a reasonable indicator of variability, although some intrinsic difference 
is expected between the two bands. In some cases, the data from different epochs show little variation and a direct 
comparison is possible. For those cases in which time-variability might be an issue, 
we will discuss the possible implications.
Line characteristics, polarisation effect properties and the measured
intrinsic polarisation, where possible, are
summarised for all objects in Table $\ref{tab:measurements}$.

In the following section, we first provide a brief discussion regarding the definition 
of observed line effects ($\S$~\ref{s_charac}), before we present the 
polarisation data of our Herbig Ae/Be objects, highlighting representative data for some individual 
objects in $\S$~\ref{s_hbe} and $\S$~\ref{s_hae}, 
while we present the remaining data with accompanying notes in the Appendix. 
We finish the section with a global comparison
between Herbig Ae and Herbig Be stars ($\S$~\ref{s_overview}). 

\subsection{Line Effect Characterisation}
\label{s_charac}

In order to characterise the observed line effects in terms 
of  ``depolarisation'' or ``$QU$ loop'', we briefly discuss 
the definitions, following the method first
used in \citet{b2}.
For depolarisation, the line effect needs to be broad compared to the
emission line. If a PA `flip' occurs , i.e. a sharp change in the profile
from above to below continuum levels, or vice versa, then this cannot be caused by depolarisation, 
but the line itself must be intrinsically polarised. 

\begin{figure*}
\centering
\includegraphics[width=50mm,  angle=-90]{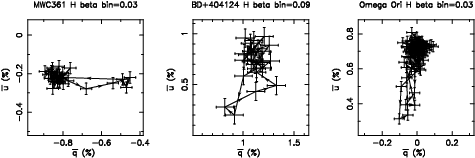}
\vfill
\vspace{2mm}
\includegraphics[width=50mm,  angle=-90]{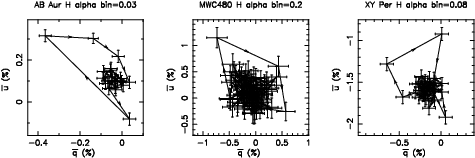}
\caption{{\it QU} data for the three Herbig Be stars with the
strongest line effect in H$\beta$ (top) and a representative sample of H$\alpha$
data of Herbig Ae stars. The Herbig Be stars show linear excursions,
while the Herbig Ae stars display loops.}
\label{F:Convince1}
\end{figure*}

In $QU$ space, the following classification can be followed: where no change 
in polarisation or PA is observed, the $QU$ loci appear as a `spot'. 
In the case of a depolarisation, a linear excursion from, and returning to, the central 
spot is observed (e.g. see the top three panels in Fig.~$\ref{F:Convince1}$), whilst in the case of a 
PA `flip', a loop is observed in $QU$ space (e.g. see the bottom three panels of Fig.~$\ref{F:Convince1}$).

\subsection{Herbig Be polarimetry at H$\gamma$, H$\beta$ and H$\alpha$}
\label{s_hbe}

Figure~$\ref{F:MWC361}$ shows first the polarisation data for MWC~361 in the top panels. This
object has the clearest depolarisation line effects in the sample and
is therefore an excellent example to start the description.

The top half of the figure shows, from left to right, the data on H$\gamma$, H$\beta$
and H$\alpha$ respectively.  The triplots show from bottom to top the
intensity spectrum, the polarisation percentage and polarisation
angle rebinned to a constant accuracy per pixel, which is shown to the
top right of each plot. Below the triplots, the {\it QU} graphs are shown.

As expected, the line emission is stronger when going to lower order
recombination lines. Indeed, H$\gamma$ emission is obviously present,
but barely exceeds the continuum, as it fills in the underlying
photospheric absorption line.  Nevertheless, even the H$\gamma$ line
shows a depolarisation similar to H$\alpha$, despite its
much smaller transition probability.  The intrinsic polarisation
measured from the line excursions is $-3^{\rm o}$ and 3$^{\rm o}$
(with errors of order 3$^{\rm o}$) respectively for the $B$ and $R$
band data. In all bands, the line-effect is best described as depolarisation, although there is a 
somewhat more complex structure in H$\alpha$ \citep[see][]{b2}.

The continuum polarisation and PA across the $R$ and $B $ bands are
quite similar, with the difference being mainly due to the slight difference
in intrinsic PA between the epochs of observation. 
The behaviour in {\it QU} space is also very similar across 
all three lines, although the blue-to-red line ratio has changed 
between the two observations. Based on the similar depolarisations in the polarisation spectra as well as the similar excursion in {\it QU} space, 
we conclude that the (de-)~polarisation mechanism responsible for the appearance at
H$\alpha$ is also acting at H$\beta$ and H$\gamma$. Therefore, a
small-scale electron scattering disc, seen close to edge-on, is 
the most plausible explanation for the phenomena observed in MWC~361.

Let us now move to the data for $\omega$~Ori, which are 
plotted in the lower panels of Fig.~$\ref{F:MWC361}$. 
This object shows depolarisation in our data
at both H$\beta$ and H$\gamma$, but it does not show any appreciable
effect in the $R$ band data of \citet{b1} taken nine years
earlier. However, \citet{b7} do observe a polarisation change at
H$\alpha$ and closer examination of the {\it QU} plots shows that the
end of the excursions at both H$\beta$ and H$\gamma$ lie near the
continuum cluster for H$\alpha$ of \citet{b1}. This corroborates their
suggestion that the decreased emission level at H$\alpha$ at the time
of observation resulted in a marked drop in the percentage of linearly
polarised scattered starlight from the star.

Also noteworthy is the fact that we see a line effect
at H$\gamma$ in the first place, although the line is in absorption.  For an
ordinary main-sequence star of the same spectral type as $\omega$~Ori,
H$\beta$ and H$\gamma$ are in absorption.  The line effect across
H$\beta$ appears to coincide only with the emission peaks, and not
with the absorption at line centre. There is some weak
emission at H$\gamma$ as well - there is just not enough to fill in
the absorption.  Never before has such a weak emission
line been observed to have a polarisation effect. This indicates that,
as long as the data is of a high enough quality, only moderate amounts
of emission are needed to cause a detectable line effect.

\subsection{Herbig Ae stars at H$\gamma$, H$\beta$ and H$\alpha$ }
\label{s_hae}

We obtained data in both the {\it B} and {\it R} band during the same
run of the Herbig Ae star MWC~480, and these are shown in
Fig. $\ref{F:MWC480}$.  MWC~480 shows no line effects in the $B$
band, but a strong line effect at H$\alpha$ is present. Contrary to
the objects discussed above, the appearance in {\it QU} space is that
of a loop.  There is evidence of emission at both H$\beta$ and
H$\gamma$, and as we showed in the previous subsection, if there would
have been a line effect, our fairly high signal-to-noise data might
have revealed it. In the case of the Herbig Be stars, the polarisation
is due to electron scattering in a disc, which does not result in a
loop in the {\it QU} diagram. The non-detection in these lines, while
we do find an intrinsic line effect at H$\alpha$ data taken nearly 
simultaneously is a further indication that the polarisation mechanism in Herbig
Ae stars may be different from that in the earlier objects.

\subsection{Brief overview of results}
\label{s_overview}

For five out of the eight Herbig Be stars that have been observed at
H$\beta$, we observe depolarisations. A comparison
with existing H$\alpha$ data shows that the intrinsic polarisation
angle is similar at both H$\alpha$ and H$\beta$. In
one case, no line-effect was detected previously at H$\alpha$, but there 
does seem to be one at shorter wavelengths now. 
This is most likely due to the intrinsic variability of the hydrogen 
recombination emission. The detection statistic, that more than half, 
but not all, of the objects are found to show a line effect is consistent 
with previous studies such as \citet{b2,b1} and \citet{b7}.

Only one of the Herbig Ae type stars displays a line effect at
H$\beta$ (MWC~120, with a spectral type of A0, the earliest Herbig Ae
star). All four objects show polarisation effects at H$\alpha$ in the form 
of a {\it QU} loop.

\section{Discussion}

\subsection{General Trends}

As has been shown by \citet{b2}, and additional data of \citet{b3}, 
the vast majority of Herbig Ae stars (10 out of 13) that display a
line-effect at H$\alpha$ show a loop in the {\it QU} diagram. 
Our new data on Herbig Ae stars indicate that the H$\alpha$ signatures 
are again of the loop variety, but we note that in the present data the 
line-effect for MWC~120 is not as clear as when it showed a loop in the 
older data of \citet{b2}. 
On the other hand, the overwhelming majority of Herbig Be objects (five
out of seven) show depolarisations at H$\alpha$. 
The only Herbig Be star that shows a loop \citep[HD~58647 in ][]{b2} has a late-B 
spectral type (B9). The H$\alpha$ line of Herbig Be stars is very strong, 
with the line-to-continuum contrasts of these objects generally exceeding 
10 \citep[see ][]{b2}, whereas the H$\alpha$ line in the Herbig Ae stars is not so 
strong, with typical line-to-continuum contrasts in the range 3-4.

\begin{table*}
\centering
\caption{H line results. Columns 2-4 contain the equivalent widths,  (with errors
below 5 per cent). Columns 5-7 indicate whether a line effect is observed, and 
columns 8-11 give the shape of the corresponding loci in $QU$ space for a given emission line \citep[defined in ][]{b2}. 
Columns 12 $\&$ 13 give the intrinsic polarisation angle, calculated by measuring Q and U for the continuum and emission line for excursion
line effect, using $\theta$($^\circ$) = $\frac{1}{2}$ arctan $(\frac{\Delta U}{\Delta Q})$.}
\begin{tabular}{@{}cccccccccccc@{}}
\hline
\centering
Object Name&\multicolumn{3}{c}{EW({\AA})}&\multicolumn{3}{c}{Line}&\multicolumn{3}{c}{QU}&\multicolumn{2}{c}{$\Theta$$_{intr}$($^\circ$) }\\
&\multicolumn{3}{c}{}&\multicolumn{3}{c}{effect?}&\multicolumn{3}{c}{behaviour}& & \\
&H$\alpha$&H$\beta$&H$\gamma$&H$\alpha$&H$\beta$&H$\gamma$&H$\alpha$&H$\beta$&H$\gamma$& $R$ & $B$  \\
\hline
MWC 166&$-$14&2.9&3.5&Y&N&N&Exc&$-$&$-$& 136 $\pm$ 4 &$-$ \\
MWC 1080$^1$&$-$&$-$14&$-$1.1&$-$&Y&Y&$-$&Smear&Smear&$-$& 158 $\pm$ 14$^*$ \\
MWC 1080$^2$&$-$100&$-$15&$-$1.4&Y&Y&Y&Smear&Smear&Smear& 165 $\pm$ 4 & 164 $\pm$ 12$^*$ \\
GU CMa&$-$14&3.8&4.8&N&N&N&$-$&$-$&$-$&$-$&$-$\\
MWC 361&$-$65&$-$2.5&3.2&Y&Y&Y&Exc&Exc&Exc& 3 $\pm$ 2 & 177 $\pm$ 4$^*$ \\
Il Cep&$-$21&3.4&4.5&N&N&N&$-$&$-$&$-$&$-$&$-$\\
BD +404124&$-$120&$-$10&$-$0.60&Y&Y&N&Exc&Exc&$-$& 36 $\pm$ 3 & 37 $\pm$ 6 \\
$\omega$ Ori&$-$2.9&1.1&2.6&N&Y&Y&$-$&Exc&Exc&$-$& 36 $\pm$ 2$^*$ \\
MWC 147&$-$65&$-$2.3&2.4&Y&Y&N&Exc/Loop?&Exc&$-$ & 168 $\pm$ 4 & 21 $\pm$ 13 \\
\hline
AB Aur&$-$32&15&16&Y&N&N&Loop&$-$&$-$&$-$&$-$\\
MWC 120&$-$20&7.6&9.5&Y&Y&Y&Loop&Loop&Loop&$-$&$-$\\
MWC 480&$-$11&16&16&Y&N&N&Loop&$-$&$-$&$-$&$-$\\
XY Per&1.9&16&13&Y&N&N&Loop&$-$&$-$&$-$&$-$\\ 
\hline 
\multicolumn{12}{l}{$^1$ 28/09/2004, $^2$ 29/09/2004, $^*$ average over H$\beta$ and H$\gamma$}\\ 
\label{tab:measurements}
\end{tabular}
\end{table*}

For the Herbig Be stars, our main finding
is that for all cases where a line
effect was found, the Herbig Be stars show the {\it same} type of line
effect at H$\beta$ and H$\gamma$ as was found earlier at H$\alpha$.
MWC~147 with a spectral type of B6 may be an exception; the results are inconclusive. 
An illustration of the differences between Herbig Be and Ae stars 
is provided in Fig.~\ref{F:Convince1}. Here we compare the H$\beta$
line in Herbig Be stars, and the H$\alpha$ line in Herbig Ae stars, as 
these lines are rather comparable in strength, yet we find a 
marked difference in the character of their line effects. 
This suggests that transition probability and optical depth do not appear to affect the type
of line effect observed, and the H$\alpha$ feature alone should be 
sufficient to decide between intrinsic line polarisation (present in the Herbig Ae/T Tauri
stars) and depolarisation (seen in the Herbig Be stars). 

These line polarisation effects may then be used as a tool to determine whether
the underlying accretion process is likely to be magnetospheric, or more
characteristic of disc accretion, as motivated in the following. 
We attribute the $QU$ loops of the Herbig Ae/T Tauri stars to magnetospheric 
accretion for a number of reasons. First of all, the line is only 
then expected to be intrinsically polarised if the photons originate {\it interior} to the 
scatterers. i.e. a compact source of line photons should be available. Secondly, the $QU$ loops 
indicate that the scatterers are situated in a rotating geometry, most likely the 
accretion disc surrounding the central star, but this is not all: 
\citet{b22} recently showed that a disc that reaches the star at its equator would 
show {\it double} PA flips rather than {\it single} ones. This indicates that the 
line polarimetry behaviour observed in the Herbig Ae/T Tauri stars is only expected 
when the accretion disc has an inner hole. Even more so, \citet{b3} constrained the sizes 
of these inner holes and found these to be compatible with magnetospheric radii.
 
The depolarisation data seen in the Herbig Be stars hint at an altogether 
different circumstellar geometry. 
First of all, the compact source of line photons is missing, whilst secondly, 
the continuum polarisation suggests the occurrence of 
electron scattering within a few stellar radii. All in all, the depolarisation data seen in 
the early Herbig Be stars are consistent with the presence of a disc, and so suggest formation by disc accretion. 
Observations in the spectral range of B4-B9 could help us understand what could 
cause this possible shift from magnetospheric accretion in
Herbig Ae stars to disc accretion in early Herbig Be stars.

\subsection{Comparison with other diagnostics}

We presented spectropolarimetric data on eight Herbig Be stars at
H$\alpha$, H$\beta$ and H$\gamma$, and we found five cases where the
behaviour could be described as ``depolarisation''. Two Herbig Be
stars show no polarisation line effects, and only one transitional
object (at spectral type B6) may show hints of a $(Q,U)$
loop. Furthermore, the observed line effects are consistent across all three
observed hydrogen lines, indicating that optical depth is not a major
factor in the observed linear polarimetry across emission lines in
Herbig stars.  This leads us to conclude that it is most likely that a
different accretion mechanism is operating in Herbig Be stars than in
Herbig Ae stars.

The fact that the role of magnetic fields may be diminishing when
going to earlier spectral types could, at first sight, be considered at
odds with the fact that up to 30 - 50\% of Herbig stars were found to be
significant emitters of X-rays \citep{b9,b10}, some of which have very early B-types. However, the higher
spatial resolution of current X-ray satellites, in particular {\sc
chandra}, can be utilised to investigate whether the X-ray photons are
intrinsic to the Herbig star, or whether they are originating from a
low-mass companion \citep[see e.g.][]{b15}. In a recent infrared (IR)
coronographic imaging and {\sc chandra} study of the extremely early
Herbig Be star MWC~297, \citet{b4} and \citet{b11} have argued that
the X-rays likely originate from a low-mass companion.  Interestingly,
an increase in the number of low-mass companions (``clustering'') with
stellar mass and spectral type amongst Herbig stars was found by
\citet{b12}, which \citet{b4} considered as yielding a higher chance
for misidentification of X-rays towards earlier type Herbig
stars. Therefore, the current status on the origin of X-rays of Herbig
Ae/Be stars suggests that T Tauri and some Herbig Ae stars may be
intrinsic X-ray emitters \citep[e.g. ][]{b19}, while early Herbig Be stars are likely not.

This picture on the X-rays from Herbig stars as a function of spectral
type can be reconciled with the linear spectropolarimetry results
presented in this paper if there is a ``fading'' of the importance of
magnetic fields at spectral types earlier than B4-B9. It is 
interesting, however, that \citet{b20} recently reported the detection of a 
significant magnetic field in the Herbig Be star MWC~361, as this is also  
the object for which \citet{b2} found a narrow PA rotation in the red wing of 
H$\alpha$, on top of an otherwise well-defined depolarisation.   

We finally mention that infrared interferometry observations 
by \citet{b13} suggest that the early Herbig Be stars (B0-B3) are discrepant 
in terms of the ``disc size - source luminosity relationship'' that works 
so well for Herbig Ae and late Herbig Be systems. 
This discrepancy could possibly be caused by the fact that 
different accretion scenarios are operating in the early Herbig Be stars 
and the Herbig Ae stars.

This appears to be in line with the line polarimetry data of 
Herbig Ae/Be stars currently available: the Herbig Ae stars show intrinsic line polarisation, 
similar to T Tauri stars, and consistent with magnetospheric accretion, while the majority of 
Herbig Be stars show depolarisation, consistent with disc accretion.

\section*{Acknowledgements}

We thank the referee for comments that improved the clarity and contents of this paper and
 Prof. Janet Drew for many fruitful discussions. JCM would like to thank Ben Davies for help with POLMAP.
The William Herschel Telescope is operated on the island of La Palma
by the Isaac Newton Group in the Spanish Observatorio del Roque de los
Muchachos of the Instituto de Astrofisica de Canarias. The allocation
of time on the WHT was awarded by PATT, the United Kingdom allocation
panel. Data analysis facilities were provided by the Starlink
Project, which is run by CCLRC on behalf of PPARC.
JCM is funded by the Particle Physics and Astronomy Research
Council of the United Kingdom. JSV acknowledges RCUK for an Academic
fellowship.

\appendix

\section{Notes on individual objects}

\subsection{Herbig Be Stars}
\label{s_A_HBe}

\begin{figure*}
\centering
\includegraphics[width=60mm, height=140mm, angle=-90]{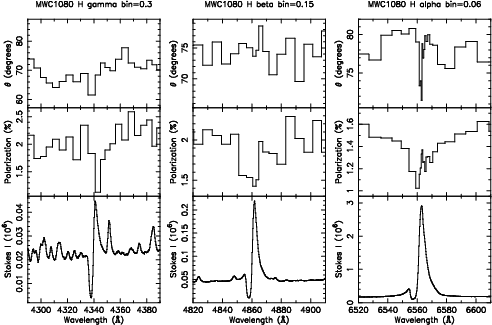}
\vfill
\vspace{2mm}
\includegraphics[width=40mm, height=140mm, angle=-90]{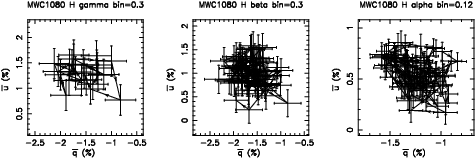}

\vspace{5mm}
\phantom{bliep} \noindent \line(1,0){475}
\vspace{5mm}

\centering
\includegraphics[width=60mm, height=140mm, angle=-90]{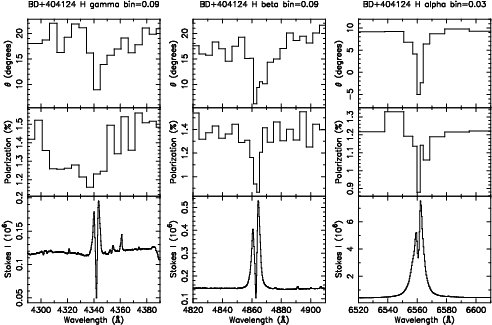}
\vfill
\vspace{2mm}
\includegraphics[width=40mm, height=140mm, angle=-90]{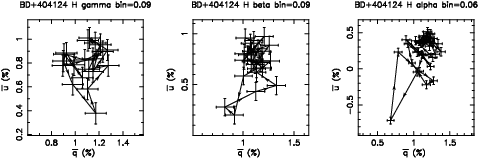}
\caption{Herbig Be Stars. Top figures: as in Fig. $\ref{F:MWC361}$ but for MWC~1080 around H$\gamma$, H$\beta$ and H$\alpha$ on 29/09/2004. The binning errors for the (Q,U) plot for H$\alpha$ and H$\beta$ are twice that of the triplot in order to provide a well-defined continuum. This data is consistent with that taken on 28/09/2004. Bottom Figures: as above but for BD~+404124 around H$\gamma$, H$\beta$ and H$\alpha$. The binning error for the (Q,U) plot for H$\alpha$ is twice that of the triplot in order to provide a well-defined continuum. The H$\alpha$ data is from 1999 while the H$\beta$ and H$\gamma$ data is from 2004.}
\label{F:MWC1080}
\label{F:BD+404124}
\end{figure*}

\begin{figure*}
\setcounter{figure}{0}
\centering
\includegraphics[width=60mm, height=140mm, angle=-90]{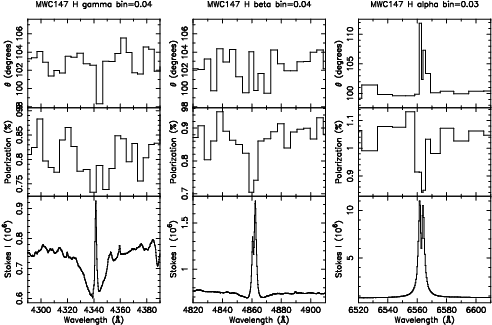}
\vfill
\vspace{2mm}
\includegraphics[width=40mm, height=140mm, angle=-90]{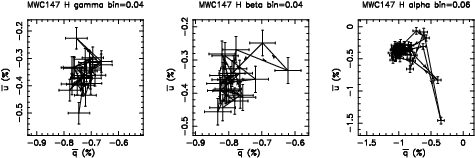}

\vspace{5mm}
\phantom{bliep} \noindent \line(1,0){475}
\vspace{5mm}

\centering
\includegraphics[width=60mm, height=140mm, angle=-90]{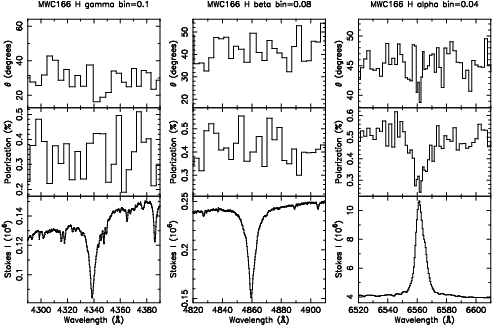}
\vfill
\vspace{2mm}
\includegraphics[width=40mm, height=140mm, angle=-90]{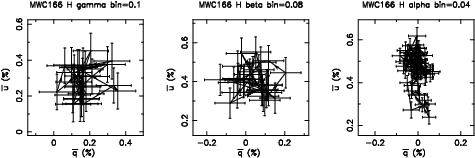}
\caption{\textbf{Cont.} Herbig Be Stars. Top figures: as in Fig. $\ref{F:MWC361}$ but for MWC~147 around H$\gamma$, H$\beta$ and H$\alpha$. The binning error for the (Q,U) plot for H$\alpha$ is twice that of the triplot in order to provide a well-defined continuum. The H$\alpha$ data is from 1999 while the H$\beta$ and H$\gamma$ data is from 2004. Bottom Figures: as above but for MWC~166 around H$\gamma$, H$\beta$ and H$\alpha$. The H$\alpha$ data is from 1996/7 while the H$\beta$ and H$\gamma$ data is from 2004.}
\label{F:MWC147}
\label{F:MWC166}
\end{figure*}

\begin{figure*}
\setcounter{figure}{1}
\centering
\includegraphics[width=60mm, height=140mm, angle=-90]{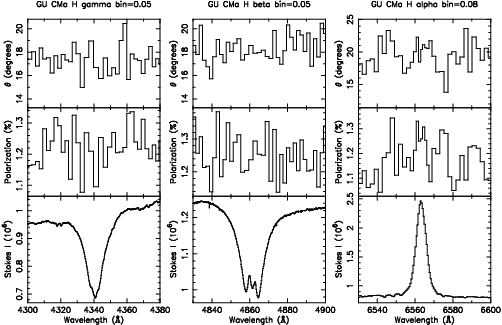}
\vfill
\vspace{2mm}
\includegraphics[width=40mm, height=140mm, angle=-90]{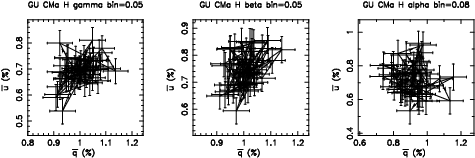}
\caption{Herbig Ae Stars. As in Fig. $\ref{F:MWC361}$ but for GU CMa around H$\gamma$, H$\beta$ and H$\alpha$. The H$\alpha$ data is from 1995 while the H$\beta$ and H$\gamma$ data is from 2004. }
\label{F:GU CMa}
\end{figure*}

\begin{figure*}
\setcounter{figure}{1}
\centering
\includegraphics[width=60mm, height=140mm, angle=-90]{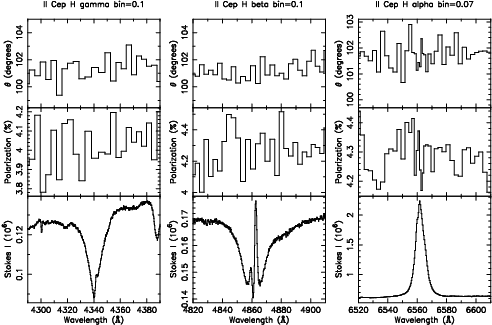}
\vfill
\vspace{2mm}
\includegraphics[width=40mm, height=140mm, angle=-90]{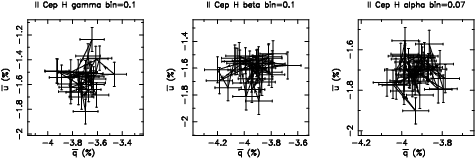}

\vspace{5mm}
\phantom{bliep} \noindent \line(1,0){475}
\vspace{5mm}

\centering
\includegraphics[width=60mm, height=140mm, angle=-90]{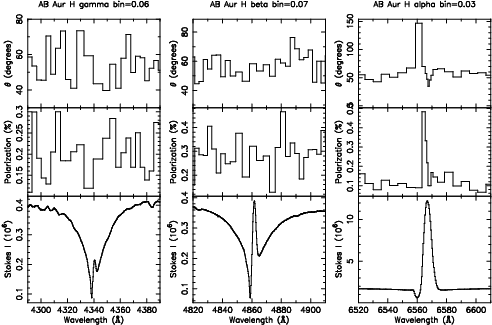}
\vfill
\vspace{2mm}
\includegraphics[width=40mm, height=140mm, angle=-90]{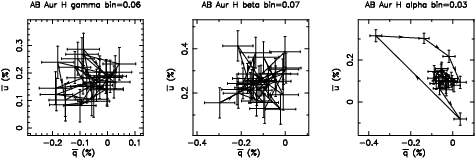}
\caption{\textbf{Cont.} Herbig Ae Stars. Top figures: as in Fig. $\ref{F:MWC361}$ but for Il Cep around H$\gamma$, H$\beta$ and H$\alpha$. The H$\alpha$ data is from 1999 while the H$\beta$ and H$\gamma$ data is from 2004. Bottom figures: as above but for AB~Aur around H$\gamma$, H$\beta$ and H$\alpha$. The H$\alpha$ data is from 1999 while the H$\beta$ and H$\gamma$ data is from 2004.}
\label{F:Il Cep}
\label{F:AB Aur}
\end{figure*}

\begin{figure*}
\setcounter{figure}{1}
\centering
\includegraphics[width=60mm, height=140mm, angle=-90]{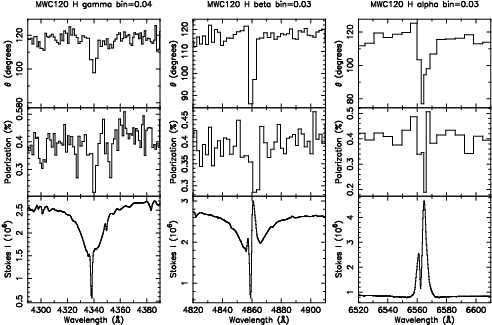}
\vfill
\vspace{2mm}
\includegraphics[width=40mm, height=140mm, angle=-90]{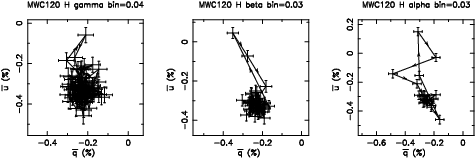}

\vspace{5mm}
\phantom{bliep} \noindent \line(1,0){475}
\vspace{5mm}

\centering
\includegraphics[width=60mm, height=140mm, angle=-90]{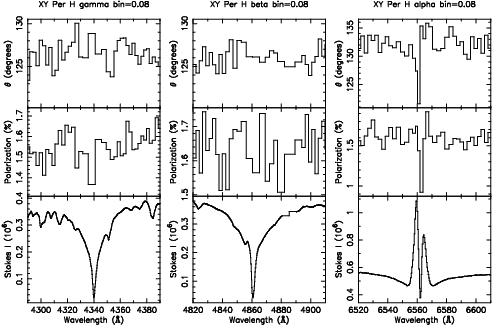}
\vfill
\vspace{2mm}
\includegraphics[width=40mm, height=140mm, angle=-90]{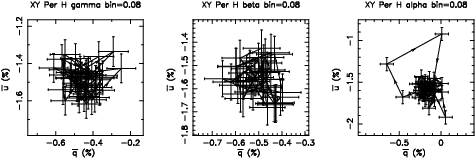}
\caption{\textbf{Cont.} Herbig Ae Stars. Top figures: as in Fig. $\ref{F:MWC361}$ but for MWC~120 around H$\gamma$, H$\beta$ and H$\alpha$. All data is from 2004. Bottom figures: as above but for XY~Per around H$\gamma$, H$\beta$ and H$\alpha$. The H$\alpha$ data is from 1999 while the H$\beta$ and H$\gamma$ data}
\label{F:MWC120}
\label{F:XY Per}
\end{figure*}

\textit{MWC~361 $\&$ MWC~1080}

These objects show line effects at all observed lines. For MWC~1080
(Fig. $\ref{F:MWC1080}$) though these are shallow and quite broad,
except in the case of H$\gamma$ on 29/04/2004. There is an increase in
$B$ band continuum polarisation between the two nights of observation,
but despite the large difference in continuum polarisation between $B$
and $R$, the intrinsic PA agrees with the data taken on the 29th. The
shape of the depolarisation effect and (Q,U) smear are similar to
those observed by \citet{b2}, although the $R$ band continuum
polarisation they record is noticeably higher. They attribute the
smear in the (Q,U) diagram to changing continuum polarisation, which
is notable between $R$ and $B$ in our observations and, to a lesser
extent, across the bands themselves. MWC~1080 would seem to be quite
variable on short timescales, and is a known binary system
\citep[][ and references therein]{b14} so will probably exhibit long
term variations as well. 

MWC361 was discussed previously in section $\ref{s_hbe}$.

~

\par\noindent\textit{BD~+404124, MWC~147, MWC~166 $\&$ $\omega$~Ori}

These objects show line effects at only some of the observed lines.
BD~+404124 shows (Fig. $\ref{F:BD+404124}$) depolarisation at H$\alpha$ and H$\beta$, with corresponding excursions in the (Q,U) plots, but 
there are no more than vague suggestions of a line effect at H$\gamma$. Both the continuum polarisation and PA increase slightly in the $B$ band 
from the $R$ band observations, possibly causing the slight change in shape of the line effect between H$\alpha$ and H$\beta$. There is, however, clear 
and quite strong emission at all three lines, so the lack of a clear effect at H$\gamma$ is most likely due to the faintness 
of BD~+404124 and may become apparent with longer exposure observations.

The first point of interest for MWC~147 is the fact that the emission lines have different shapes in the $R$ and $B$ bands 
(Fig. $\ref{F:MWC147}$). H$\alpha$ (observed 18/12/1999) is double peaked with a V/R peak ratio $>$ 1, i.e. V~$>$~R,  while H$\beta$ (observed 28/09/2004) is 
double peaked with V $<$ R. H$\gamma$ appears single peaked, with the emission slightly displaced to the R side of the absorption feature. 
This difference between H$\alpha$ and H$\beta$ is probably due to time variability. 
There are definitely line effects at H$\alpha$ and H$\beta$, with little sign of effects at H$\gamma$, but it is unclear 
if the effects are depolarisation effects, or QU loops. 
This makes it difficult to say if MWC~147 is more likely accreting via magnetospheric accretion or via disc accretion.

For MWC~166, a depolarisation is observed at H$\alpha$ while no line effects are observed in the $B$ band 
(Fig. $\ref{F:MWC166}$). The continuum polarisation in the $B$ band is lower than at $R$, to the point where the continuum 
polarisation in $B$ is near the bottom of the depolarisation at H$\alpha$. In addition, there are only 
slight indications of emission at H$\beta$ or H$\gamma$. 
Overall, it is to be expected that we do not see line effects for MWC~166 in $B$, 
with the differences possibly caused by variations in the object between the two observations.

$\omega$~Ori was discussed previously in section $\ref{s_hbe}$.

~

\par\noindent\textit{GU CMa $\&$ Il Cep}

Neither GU CMa (Fig. $\ref{F:GU CMa}$), nor Il Cep (Fig. $\ref{F:Il
Cep}$) show any sign of line effects in either the $B$ or $R$
band. Although the $R$ band data was taken at a different epoch to 
the $B$ band data, the continuum polarisation percentage and PA 
agree, which suggests little change in the circumstellar medium around the 
object between these observations. The high level of
continuum polarisation of the objects, $\sim$1.2$\%$ for GU~CMa and
$\sim$4.2$\%$ for IL~Cep, and the high quality of the data means
that it is unlikely that these objects exhibit undetected line
effects, and their electron scattering discs are expected to be relatively pole-on.

\subsection{Herbig Ae Stars}
\label{s_A_HAe}

\textit{AB~Aur, MWC~120, MWC~480 $\&$ XY~Per}

These objects have line effects at only some of the observed lines.  AB
Aur shows clear line effects in the H$\alpha$ data of \citet{b2} but
we observe no line effects in our $B$ band data (Fig. $\ref{F:AB
Aur}$). The continuum polarisation has increased considerably,
although \citet{b3} see line effects at H$\alpha$ in Dec 2003 with a
level of continuum polarisation in $R$ similar to ours.  Therefore,
while the polarisation level around AB~Aur has increased, it should
not have prevented detection of the line effect. There are signs of
emission in the H$\beta$ and H$\gamma$ profiles, but it is possible
that it is simply not high enough for an A type star, which tend to
have broader and deeper absorption lines and lower continuum emission
at these wavelengths than B type stars.

MWC~120 shows the suggestion of a line effect at H$\gamma$ and clear line effects at the other two 
observed lines (Fig. $\ref{F:MWC120}$). The V/R ratio appears to be similar in both H$\beta$ and 
H$\alpha$, and there are signs of some emission at H$\gamma$. The data is of quite a high signal-to-noise, 
so longer exposure may not improve the definition of the possible effect at H$\gamma$.

XY~Per shows no sign of line effects in $B$ but does show line effects at H$\alpha$ (Fig. $\ref{F:XY Per}$). 
The continuum polarisation is consistent between the two observed bands, and there are few signs of emission 
in the $B$ band line profiles, suggesting that there are too few photons for a signature to be observed.

MWC~480 was discussed previously in section $\ref{s_hae}$

\label{lastpage}


\begin{thebibliography}{99}

\bibitem[\protect\citeauthoryear{Baines et al.}{2006}]{b14} Baines D., Oudmaijer R.D., Porter J.M., Pozzo M., 2006, MNRAS, 367, 737
\bibitem[\protect\citeauthoryear{Damiani et al.}{1994}]{b9} Damiani F., Micela G., Sciortino S., Harnden F.R., 1994, ApJ, 436, 807
\bibitem[\protect\citeauthoryear{Damiani et al.}{2006}]{b11} Damiani F., Micela G., Sciortino S., 2006, A\&A, 447, 1041
\bibitem[\protect\citeauthoryear{Hubrig et al.}{2006}]{b17} Hubrig S., Yudin R.V., Sch\"oller M., Pogodin M.A., 2006, A\&A, 446, 1089
\bibitem[\protect\citeauthoryear{Johns-Krull et al.}{1999}]{b6} Johns-Krull C.M., Valenti J.A., Hatzes A.P., Kanaan A., 1999, ApJ, 510, L41
\bibitem[\protect\citeauthoryear{Monnier at al.}{2005}]{b13} Monnier J.D., Millan-Gabet R., Billmeier R. et al., 2005, ApJ, 624, 832
\bibitem[\protect\citeauthoryear{Monet et al.}{2003}]{b21} Monet D.G., Levine S.E., Canzian B. et al., 2003, AJ, 125, 984
\bibitem[\protect\citeauthoryear{Natta et al.}{2001}]{b5} Natta A., Prusti T., Neri R., Wooden D., Grinin V.P., Mannings V., 2001, A\&A, 371, 186
\bibitem[\protect\citeauthoryear{Oudmaijer ~\&~ Drew}{1999}]{b1} Oudmaijer R.D., Drew J.E., 1999, MNRAS, 305, 166
\bibitem[\protect\citeauthoryear{Poeckert ~\&~ Marlborough}{1975}]{b7} Poeckert R., Marlborough J.M., 1976, ApJ, 206, 182
\bibitem[\protect\citeauthoryear{Stelzer et al.}{2006}]{b15} Stelzer B., Micela G., Hamaguchi K., Schmitt J.H.M.M., 2006, A\&A, 457, 223
\bibitem[\protect\citeauthoryear{Swartz et al.}{2005}]{b19} Swartz D.A., Drake J.J., Elsner R.F. et al., 2005, ApJ, 628, 811
\bibitem[\protect\citeauthoryear{Symington et al.}{2005}]{b8} Symington N.H., Harries T.J., Kurosawa, R., Naylor T., 2005, MNRAS, 358, 977
\bibitem[\protect\citeauthoryear{Testi et al.}{1998}]{b12} Testi L., Palla F., Natta A., 1998, A\&AS, 133, 81
\bibitem[\protect\citeauthoryear{Th{\'e} et al.}{1994}]{b18} Th{\'e} P.S., de Winter D., Perez M.R., 1994, A\&AS, 104, 315
\bibitem[\protect\citeauthoryear{Vink et al.}{2002}]{b2} Vink J.S., Drew J.E., Harries T.J., Oudmaijer R.D., 2002, MNRAS, 337, 356
\bibitem[\protect\citeauthoryear{Vink et al.}{2005a}]{b3} Vink J.S., Drew J.E., Harries T.J., Oudmaijer R.D., Unruh Y., 2005a, MNRAS, 359, 1049
\bibitem[\protect\citeauthoryear{Vink et al.}{2005b}]{b22} Vink J.S., Harries T.J., Drew J.E., 2005b, A\&A, 430, 213
\bibitem[\protect\citeauthoryear{Vink et al.}{2005c}]{b4} Vink, J.S., O'Neill P.M., Els S.G., Drew J.E., 2005c, A\&A, 438, L21
\bibitem[\protect\citeauthoryear{Wade et al.}{2005}]{b16} Wade G.A., Drouin D., Bagnulo S. et al., 2005, A\&A, 442L, 31
\bibitem[\protect\citeauthoryear{Wade et al.}{2006}]{b20} Wade G.A., Drouin D., Bagnulo S., et al., 2006, astro-ph/0601624
\bibitem[\protect\citeauthoryear{Zinnecker ~\&~ Preibisch}{1994}]{b10} Zinnecker H., Preibisch Th., 1994, A\&A, 292, 152


\end{thebibliography}
\end{document}